\documentclass[
reprint,
superscriptaddress,
 amsmath,amssymb,
 aps,
prl,
]{revtex4-2}
\usepackage{xcolor}% text color
\usepackage{graphicx}% Include figure files
\usepackage{dcolumn}% Align table columns on decimal point
\usepackage{bm}% bold math
\usepackage{subeqnarray}
\usepackage[normalem]{ulem}
\usepackage{hyperref}
\hypersetup{
    colorlinks=true,
    linkcolor=blue,
    filecolor=blue,      
    urlcolor=blue,
    citecolor=blue
    }
\begin{document}

\title{Solubility enhanced surfactant-induced flow in air-liquid-air sheets}

\author{Jun Eshima}
\email{jeshima@princeton.edu}
\affiliation{Department of Mechanical and Aerospace Engineering, Princeton University, Princeton, New Jersey 08544, USA}
\author{Tristan Aurégan}
\affiliation{Department of Mechanical and Aerospace Engineering, Princeton University, Princeton, New Jersey 08544, USA}
\author{Emmanuel Villermaux}
\affiliation{Aix Marseille Université, CNRS, Centrale Marseille, IRPHE UMR 7342, 13384 Marseille, France}
\affiliation{Institut Universitaire de France, 75005 Paris, France}
\author{Howard A. Stone}
\email{hastone@princeton.edu}
\affiliation{Department of Mechanical and Aerospace Engineering, Princeton University, Princeton, New Jersey 08544, USA}
\author{Luc Deike}
\email{ldeike@princeton.edu}
\affiliation{Department of Mechanical and Aerospace Engineering, Princeton University, Princeton, New Jersey 08544, USA}
\affiliation{High Meadows Environmental Institute, Princeton University, Princeton, New Jersey 08544, USA}

\begin{abstract}
Liquid interfaces appear throughout nature and engineering and are typically contaminated by surface active agents (surfactants), which are characterized by a wide range of solubility.
We demonstrate that solubility enhances by an order of magnitude surfactant-induced flow in air-liquid-air films, in contrast to previously studied geometries where solubility dampens the flow. The enhancement is described by a single parameter comparing the depletion length to the film thickness. Our experiments are well described by an asymptotic theory of the Navier-Stokes equations, including surfactant kinetics.
\end{abstract}

\maketitle

Interfaces influence dynamics in the environment, health, and industrial processes. 
For example, sea spray aerosols from bursting bubbles strongly affect ocean-atmosphere mass exchange \cite{Veron15,Deike22}.
Exhaled aerosols transmit pathogens \cite{Bourouiba21}, while thin liquid films line the lungs \cite{Possmayer23} and eyes \cite{Braun12}. 
In industry, emulsions, coatings, and foams exhibit intricate interfacial flow dynamics.

Furthermore, most liquid interfaces are contaminated. A common type is surface active agents or surfactants: molecules of both biological and anthropogenic origins that adsorb to the interface and lower surface tension. These diverse origins yield surfactants with distinct chemical properties governing their physicochemical behavior \cite{Fuller12,Manikantan20}. 
For example, the solubility (adsorption equilibrium constants) of biological surfactants, including polysaccharides, lipids, proteins, and humics, in the ocean can vary by several orders of magnitude \cite{Burrows14}. 
The formation of sea spray aerosols involves several surfactant-dependent processes, including wave breaking \cite{Liu03, Erinin23}, bubble scavenging during flotation \cite{Blanchard89}, turbulent bubble breakup \cite{Soligo19,Wu26}, bubble lifetime \cite{Poulain18}, and bursting dynamics \cite{Constante21,Constante22}. These processes can be modulated by surfactant properties, of which solubility is an important example, contributing to the variation in relative abundance of compounds between ocean surface water and sea spray aerosols \cite{Burrows14}. Such complexity remains a major source of uncertainty in environmental models. 

Surfactant concentration gradients produce surface tension gradients, which drive so-called \textit{Marangoni} flow. 
A commonly studied configuration is the localised deposition of surfactant onto a clean air-liquid interface, generating outward spreading flow. This configuration has been studied for close to two centuries \cite{Thomson1855,Matar09}, primarily on a deep liquid subphase or on thin air-liquid-solid films. In both configurations, surfactant solubility, a chemical property, reduces the spreading rate \cite{Joos85,Banos25,Halpern92,Jensen93}, a flow property, as surface surfactants desorb into the bulk, leading to a weaker Marangoni stress due to a lower surface surfactant gradient \cite{Banos25,Halpern92,Jensen93}. 

In contrast, we focus on surfactant deposition on an air-liquid-air film \cite{Bowen13,Neel18,Kitavtsev18,Motaghian19,Eshima25_PRL,Eshima25_JFM}, which is less studied despite its ubiquity. 
For example, localised Marangoni stress is a proposed  mechanism for rupture of a bubble cap (an air-liquid-air film)  \cite{Neel18, Poulain18}, thereby influencing bubble lifetime, and in turn, oceanic aerosol production. More generally, air-liquid-air films arise in foam and soap films \cite{Langevin20,Cantat26}, bubble coalescence \cite{Chan11}, and the breakup of rain drops \cite{Villermaux09}. Understanding when and how such films thin and rupture is therefore a recurring question. 

\begin{figure}
\includegraphics[width=0.49\textwidth]{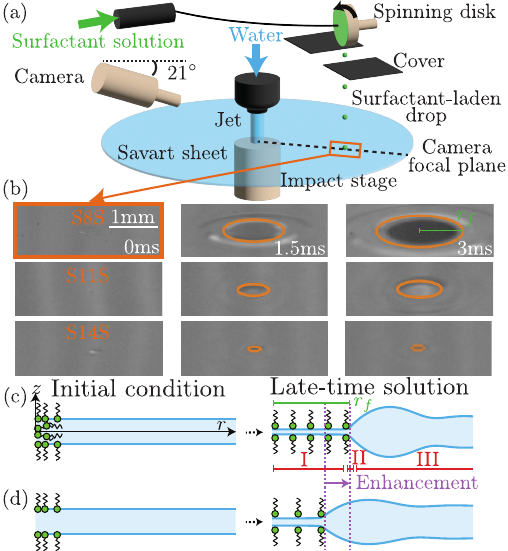} 
\caption{\label{fig:setup} Summary of the surfactant deposition experiment and solubility-driven enhancement of front propagation and film thinning. (a) Experimental setup for surfactant deposition onto an air-liquid-air (Savart) sheet, produced by a jet impacting a stage. Monodisperse surfactant-laden drops are produced by a spinning disk atomiser. (b) Sample dynamics of surfactant-induced front propagation with radial distance $r_f$. Frames are centered on the deposition pattern. Rows are surfactants of decreasing solubility (top to bottom): S8S, S11S, S14S, at the half concentration case (see table \ref{tab:surfactant_values}), corresponding to static surface tension deficit $\Delta \Sigma_0\approx 2-3$ mN/m. Columns show time evolution. Axisymmetric patterns appear elliptic due to the camera angle of $\approx 21^{\circ}$ to the horizontal. The orange curves overlaid on the experimental frames are the theoretical predictions without any fitting parameters (ratio of minor to major axes of the overlaid ellipse is set as $\sin(21^{\circ})$). \cite{SM_letter} See Supplemental Material for video versions. (c) Schematic film evolution. (d) Analogous to (c) but without bulk surfactants. The purple arrow shows the enhancement of surfactant front propagation and hence the thinning. Diagrams in (a,c,d) are not to scale.}
\end{figure}

Specifically, we consider a surfactant-laden drop deposited onto a Savart sheet \cite{Marmottant00,Clanet02,Neel18} (Fig.~\ref{fig:setup}(a)).
The deposition leads to a localised region of low surface tension, driving an outward radial flow and sheet thinning; Fig.~\ref{fig:setup}(b) shows spreading patterns with front radius $r_f(t)$.

Surfactant deposition on thin air-liquid-air films of low viscosity liquids (such as water) is characterized by the deposition length scale (along the film) $\mathcal{L}$, the initial uniform film thickness $h_i$, and a timescale $\mathcal{T}$ obtained by balancing inertia $\rho h_i \mathcal{L}\mathcal{T}^{-2}$ (where $\rho$ is the liquid density) and Marangoni stresses $\Delta \Sigma_0 \mathcal{L}^{-1}$ arising from the surfactant-induced surface tension deficit $\Delta \Sigma_0$: $\mathcal{T} = (\rho h_i \mathcal{L}^{2}/(\Delta \Sigma_0))^{1/2}$ \cite{Neel18,Eshima25_PRL,Eshima25_JFM}.

Taking typical values of $\mathcal{L} = \textit{O}(100 \text{ \textmu m})$, $\Delta \Sigma_0 = \textit{O}(1 \text{ mN/m})$, and $h_i = \textit{O}(10 \text{ \textmu m})$, the magnitude of $\mathcal{T}$ is $\textit{O}(100 \text{ \textmu s})$. 
In the late-time spreading limit where time $t \gg \mathcal{T}$, experiments \cite{Neel18} with ethanol drops showed that the surfactant front $r_f(t)$ propagates with $r_f(t)/\mathcal{L} \sim (t/\mathcal{T})^{1/2}$. 
In the thin-film limit ($h_i \ll \mathcal{L}$) with insoluble surfactants, the Navier-Stokes equations simplify, and the late-time $t^{1/2}$ propagation was derived theoretically \cite{Eshima25_PRL,Eshima25_JFM}, accounting for inertia, Marangoni, capillary, and viscous extensional stresses with surfactant transport. 

The insoluble surfactant dynamics are described by two nondimensional numbers (see end matter): the Marangoni number $\mathcal{M}$ (ratio of Marangoni to capillary stresses), and Reynolds number $\textit{Re}$ (ratio of inertia to viscous extensional stresses). 
For typical values of $\mathcal{L}$ and $\mathcal{T}$ as above, diffusion along the sheet is not considered: the Péclet number $\mathcal{L}^2/(D\mathcal{T})=\textit{O}(10^6) \gg 1$ since the typical diffusion coefficient $D = \textit{O}(10^{-10} \text{ m$^{2}$/s})$. 

Here, we show experimentally and theoretically that surfactant solubility strongly enhances the front propagation, in contrast to the reduction observed on air-liquid-solid films and deep liquid subphases. We show that the $t^{1/2}$ scaling for $r_f(t)$ still holds,
\begin{equation}
    r_f(t) =\eta_f \mathcal{L} \left(\frac{t}{\mathcal{T}}\right)^{1/2}=\eta_f \left(\frac{\Delta \Sigma_0 \mathcal{L}^2}{\rho h_i}\right)^{\frac{1}{4}}t^{\frac{1}{2}},\label{eq:front_prefac}
\end{equation}
demonstrate experimentally that solubility increases the prefactor $\eta_f$, and derive a theoretical prediction for it. The overlaid curves on Fig.~\ref{fig:setup}(b) are the theoretical predictions (to be derived) without fitting parameters.

The enhancement arises since in an air-liquid-air film, the surfactant-containing region thins as the front $r_f(t)$ propagates, which pushes bulk surfactant to adsorb onto the surface, strengthening the Marangoni stress (Fig.~\ref{fig:setup}(c,d)).
This push can be understood by noting that incompressible stretching of a material element increases its surface-area-to-volume ratio: stretching acts to dilute the surface surfactants, while acting to retain bulk surfactant concentration, and surfactants therefore adsorb onto the surface.
Thus, the bulk fluid acts as a reservoir that replenishes the surface with surfactants. 
Bulk replenishment of the surface is central to many surfactant phenomena \cite{Manikantan20, Bussonniere21, Stebe91, Fernandez26, Guillemot26}. 
Equivalently, the interface acts as a sink for bulk surfactants.

The late-time propagation for finitely soluble surfactants has the same spatial structure (Fig.~\ref{fig:setup}(c)) as for insoluble surfactants, which consists of a propagating surfactant front (region II), a thin region with spatially uniform surfactant concentration behind the surfactant front (region I), and a region ahead of the surfactant front without surfactants (region III). 
This structure is physically expected since inertial surfactant flow for air-liquid-air films mathematically maps onto the compressible gas equations upon identifying surface surfactant concentration with (negative) gas pressure and film thickness with gas density \cite{Chomaz01,Kim17, Eshima25_JFM} (see end matter). 
The surfactant front propagation therefore has analogies to classical compressible gas shock propagation.  

To quantify enhancement via solubility, we account for surfactant physicochemistry. We consider low bulk and surface surfactant concentrations where adsorption and desorption fluxes are linear.
Then, the surface surfactant concentration $\Gamma_0$ at kinetic equilibrium with the bulk surfactant concentration $c_0$ satisfies 
\begin{equation}
    k_d\Gamma_0=k_a c_0 ,
\end{equation}
where $k_a$ and $k_d$ are constants. Similarly, at dilute concentrations, surface tension $\sigma(\Gamma)$ varies linearly with surface surfactant concentration $\Gamma$. Hence the static surface tension deficit $\Delta \Sigma_0$ for $c_0$ is 
\begin{equation}
    \Delta \Sigma_0 = -\left.\frac{d\sigma}{d\Gamma}\right|_{\Gamma=0} k_a k_d^{-1} c_0.\label{eq:static_surface_tension}
\end{equation}

A natural lengthscale $k_ak_d^{-1}$ appears, which is commonly referred to as the depletion length \cite{Manikantan20} or equilibrium adsorption constant. Estimates for $k_ak_d^{-1}$ are available from published equilibrium measurements (table \ref{tab:surfactant_values}). Smaller $k_ak_d^{-1}$ means that a higher $c_0$ is required to obtain the same $\Gamma_0$. The nondimensional parameter comparing surface and bulk  concentrations at equilibrium is 
\begin{equation}
    \Lambda_d = \frac{k_a}{k_d h_i} = \frac{\Gamma_0}{c_0h_i},\label{eq:Lambda_d}
\end{equation}
where $h_i$ is the initial thickness of the liquid sheet onto which surfactants are deposited. 

Our experimental setup is as follows. We use sodium alkyl sulfates (CH$_3$(CH$_2$)$_{n-1}$OSO$_3$Na), which are well-documented anionic surfactants \cite{Chang95, Varga07} with a hydrophobic hydrocarbon tail and hydrophilic sodium sulfate head, henceforth labeled S$n$S. 
The choice allows us to systematically explore surfactant solubility since a longer hydrocarbon tail length $n$ leads to lower solubility. We use S$n$S for $n = 8,\cdots, 14$ as purchased (Thermo Fisher, Fisher Scientific, Neta Scientific), which are used in many cleaning and personal care products. Furthermore, S$n$S are often used as model surfactants, e.g., as proxies for lipids found in the ocean \cite{Burrows14}. In particular, S12S is known as SDS (sodium dodecyl sulfate). We use solutions of S$n$S dissolved in deionised (DI) water. 
As the base case, the bulk surfactant concentration $c_0$ of each S$n$S solution is set (table \ref{tab:surfactant_values}) so that the static surface tension of the solution is $67 \pm 1$ mN/m (Wilhelmy plate method), giving a deficit $\Delta \Sigma_0 \approx 5$ mN/m relative to pure DI water ($72$ mN/m).
The concentrations $c_0$ are $\approx 10$\% of the critical micelle concentration for each S$n$S. Experiments at double and half concentrations are also performed to check sensitivity to $c_0$ and the associated $\Delta \Sigma_0$.

\begin{table}[t]
\centering
\begin{tabular}{c ccccccc}
\hline\hline
 & S8S & S9S & S10S & S11S & S12S & S13S & S14S \\
\hline
$k_ak_d^{-1}$ [\textmu m] & 0.07 & 0.15 & 0.3 & 0.5 & 1.0 & 2.0 & 3.5 \\
$c_0$ [mM] & 13 & 6.0 & 2.6 & 1.5 & 0.84 & 0.44 & 0.29 \\
\hline\hline
\end{tabular}
\caption{Depletion lengths $k_ak_d^{-1}$ for sodium alkyl sulfates obtained by fitting the curve $k_d \Gamma_0 = k_a c_0$ to figure 4b of \cite{Varga07}. We allow for $\pm 30\%$ uncertainty range of $k_ak_d^{-1}$. The base case concentration $c_0$ of the sodium alkyl sulfate solutions with measured surface tension of $67\pm 1$ mN/m is shown. Experiments are also performed at double and half the base case concentration.}
\label{tab:surfactant_values}
\end{table}

The surfactant-laden drops are produced by placing the surfactant solution on a fast spinning disk enabling controlled production of small monodisperse drops in all directions parallel to the disk plane \cite{Davies84,Bocanegra15}; a cover selects the deposition location. Such control is necessary for quantitative theoretical comparisons.
By separately imaging the disk, we estimate the mean drop radius to be $r_d \approx 22\pm 2$ \textmu m for a disk of radius $1$ cm rotating at angular frequency $\approx 2000$ rad s$^{-1}$. Since the operating conditions are held constant, this mean drop radius is taken to apply across all experiments. 

The air-liquid-air sheet, referred to as a Savart sheet, is formed by a (tap) water jet of diameter $d_j \approx 3.3$ mm with velocity $v_j \approx 4$ m s$^{-1}$ striking a horizontal impact stage, where an adjustable sleeve allows the sheet to be produced at a right-angle to the jet \cite{Marmottant00, Clanet02}. 
A nozzle with a large contraction ratio keeps the jet, and hence the sheet, laminar at high velocities.
The camera field-of-view, onto which the surfactant deposition is aimed, is located at a radial distance $r_s \approx 5-8$ cm from the jet. The Savart sheet thickness formula \cite{Marmottant00,Clanet02} gives $ 0.13d_j^2/r_s = 20-30$ \textmu m, which we checked by interferometry. 

By imaging at 22000 fps (Phantom v2012 paired with a Nikon Micro 200 mm and 36 mm extension tube, giving $\approx 20$ \textmu m/pixel resolution), we track the evolution of the front position $r_f(t)$, which appears in the image as the major radius of an ellipse due to the imaging angle. 
We saved instances where the drop deposition was visible in the frame, recording until either rupture or advection out of the frame.
Sample images of the experiment are shown in Fig.~\ref{fig:setup}(b). Here, for S8S in Fig.~\ref{fig:setup}(b), the black regions are not holes but extremely thin films. Once a hole forms, Taylor-Culick-type retraction \cite{Taylor59,Culick60,Mcentee69,Savva09,Guillemot26} is rapid and readily visible.  

\begin{figure}
\includegraphics[width=0.49\textwidth]{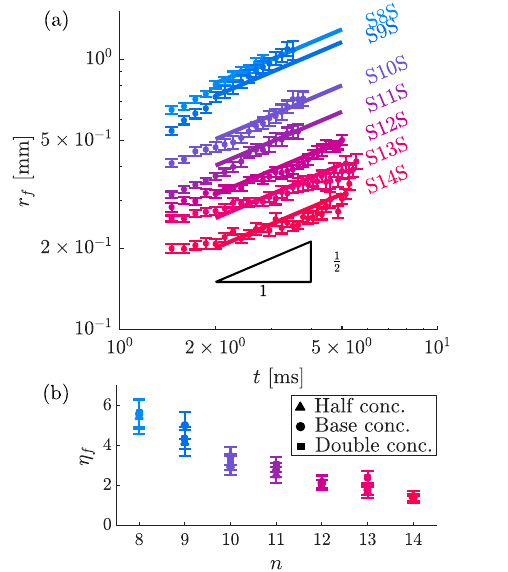} 
\caption{\label{fig:exp_analysis} Experimental data of the enhancement of front propagation via surfactant solubility. (a) Surfactant front radius $r_f$ against time $t$ for each surfactant (S8S to S14S shown as blue to red) for the base case concentrations shown in table \ref{tab:surfactant_values} where surface tension deficit $\approx 5$ mN/m. The error bars show the mean $\pm$ 1 standard error. The solid lines show the best linear least-squares fits to a $t^{1/2}$ power law. 
(b) The nondimensional prefactor $\eta_f$ (\ref{eq:front_prefac}, \ref{eq:explicit_L}) against $n$, the number of carbons in S$n$S. Here, $r_d = 22$ \textmu m is the drop radius, $\rho = 997$ kg m$^{-3}$  the water density, $h_i = 25$ \textmu m the initial thickness of the liquid sheet, and $\Delta \Sigma_0 = 2.5,5,10$ mN/m the static surface tension deficit of the surfactant solution for half (triangle), base (dot), and double (square) concentrations respectively. The error bars are from the uncertainty of the values $\Delta \Sigma_0$ ($\pm 1$ mN/m), $r_d$ ($\pm 2$ \textmu m), and $h_i$ ($\pm 5$ \textmu m).}
\end{figure}

The temporal evolution of the front radius for the base concentration case ($\approx 5$ mN/m surface tension deficit) is shown in Fig.~\ref{fig:exp_analysis}(a). Each case is repeated 30 times. The mean and standard error are shown for $t>1.5$ ms. At earlier times, the spreading front is often not clearly visible; our focus is on the late-time propagation. 
The solid lines show linear least-squares fits to a $t^{1/2}$ power law, in good agreement with the scaling predicted for insoluble surfactant deposition. 
Treating the exponent as free parameter shows values close to $1/2$ (e.g., $0.5 \pm 0.15$ for the data shown in Fig.~\ref{fig:exp_analysis}(a)). We do not analyze these deviations further, attributing them to experimental uncertainties and transient effects since the $t^{1/2}$ scaling only holds at late times $t \gg \mathcal{T}$ while our observational window is $t/\mathcal{T}= \textit{O}(10-100)$ for $\mathcal{T} = \textit{O}(100 \text{ \textmu s})$.

The fitted prefactor $\eta_f$ is defined from (\ref{eq:front_prefac}) by setting
\begin{equation}
    \mathcal{L} = \sqrt{\frac{4 r_d^3}{3h_i}},\label{eq:explicit_L}
\end{equation}
where $\mathcal{L}$ is obtained by equating the characteristic sheet volume $\pi \mathcal{L}^2h_i$ to the drop volume $(4/3)\pi r_d^3$.
Substituting (\ref{eq:explicit_L}) into (\ref{eq:front_prefac}), we have $r_f = \eta_f [(4\Delta \Sigma_0r_d^3)/(3\rho h_i^2)]^{1/4}t^{1/2}$.
The plot of $\eta_f$ against $n$ is shown in Fig.~\ref{fig:exp_analysis}(b). 
Two additional concentrations are shown, where the concentration is doubled and halved respectively, i.e., $\Delta \Sigma_0\approx 10$ mN/m and $\Delta \Sigma_0\approx 2.5$ mN/m, where data points are obtained as in Fig.~\ref{fig:exp_analysis}(a). 
Fig.~\ref{fig:exp_analysis}(b) shows that for fixed $\Delta \Sigma_0$, $\eta_f$ increases as $n$ decreases, corresponding to higher solubility; 
the same enhancement is also seen in Fig.~\ref{fig:setup}(b). 
For fixed $n$, all $\eta_f$ collapse across $\Delta \Sigma_0$, meaning that larger surface tension deficits result in faster front propagation, consistent with (\ref{eq:front_prefac}).
Overall, varying $n$ over the range considered changes the front growth by factor $\approx4$, suggesting that solubility is a key parameter in characterizing the dynamics.

Now, we propose a theoretical model for our experimental measurements.
First, we consider the sheet. Since the field-of-view is small, 
the sheet is locally flat and uniformly translating; background stretching is weak (see end matter), consistent with the axisymmetric patterns experimentally observed in Fig.~\ref{fig:setup}(b).
Additionally, forming a Savart sheet creates large surface area and hence the surface of the sheet is devoid of additional or parasitic surface active compounds \cite{Marmottant00,Neel18}. Next, the drop retains the original solution concentration $c_0$ and hence the total surfactant in the deposited drop is $(4/3) \pi r_d^3 c_0$. 
Since the deposited drops are small enough that coalescence occurs rapidly relative to our experimental measurement timescale (see end matter), the drop impact and coalescence dynamics can be neglected.
Thus, we consider the model problem of axisymmetric surfactant deposition onto a clean flat film of uniform thickness $h_i$ otherwise at rest.

At late times, the region containing the surfactants is thin and the surfactants are pushed onto the surface due to increased surface-area-to-volume ratio. 
Since the surfactants are mostly on the surface, the late-time similarity solution that describes the film dynamics here is the same as that of insoluble surfactants with the same total amount of surfactants. 
Then, the theoretical prediction for the enhanced soluble surfactant front growth is derived by modifying the late-time similarity solution for insoluble surfactants deposited onto a clean flat film otherwise at rest  \cite{Eshima25_PRL,Eshima25_JFM}.
Since the characteristic surface surfactant concentration is $h_ic_0/2$ (distributing $h_i c_0$ between top and bottom surfaces), there is an enhancement factor  $h_ic_0/(2\Gamma_0) = (2\Lambda_d)^{-1}$ over $\Gamma_0$. This gives (see end matter) an effective surface tension deficit $\Delta \Sigma_{\text{eff}}$: 
\begin{equation}
    \Delta \Sigma_{\text{eff}} = \frac{\Delta \Sigma_0}{2\Lambda_d},\label{eq:eff_surf_tension}
\end{equation}
and
\begin{equation}
    r_f = \eta_f^{\text{insol}} \left(\frac{4\Delta \Sigma_{\text{eff}} r_d^3}{3\rho h_i^2}\right)^{\frac{1}{4}}t^{\frac{1}{2}}, \label{eq:r_f_eff}
\end{equation}
where $\eta_f^{\text{insol}} = \eta_f^{\text{insol}}(\mathcal{M}, \textit{Re})$ is a function of $\mathcal{M}$ and $\textit{Re}$ which weakly depends on $\Lambda_d$ and is obtained numerically from the similarity solution to the  ordinary differential equations for insoluble surfactant deposition.
Then, comparing (\ref{eq:eff_surf_tension}, \ref{eq:r_f_eff}) with (\ref{eq:front_prefac}, \ref{eq:explicit_L}) gives 
\begin{equation}
    \eta_f = \eta_f^{\text{insol}}(2\Lambda_d)^{-\frac{1}{4}}.\label{eq:front_prefac_prediction}
\end{equation}

In particular, it can be shown analytically that for $\Lambda_d \ll 1$, the function $\eta_f^{\text{insol}}$ tends to a constant: $\lim_{\Lambda_d\rightarrow 0} \eta_f^{\text{insol}}(\mathcal{M},\textit{Re})=2$ \cite{Eshima26_long_paper}. Thus, for $\Lambda_d \ll 1$,
\begin{equation}
    \eta_f =  2^{\frac{3}{4}}\Lambda_d^{-\frac{1}{4}},\label{eq:front_asymptote}
\end{equation}
which is a scaling law for $\Lambda_d$ with an analytical prefactor.

\begin{figure}
\includegraphics[width=0.49\textwidth]{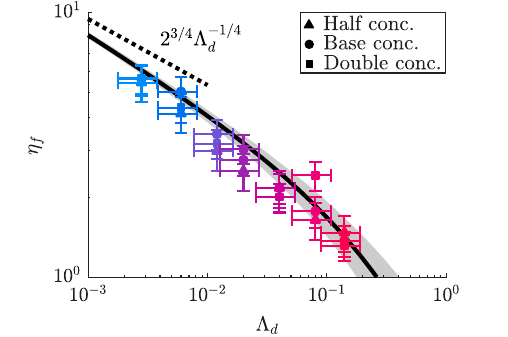} 
\caption{\label{fig:theory_comp} Comparison of the experimental data and theoretical prediction of the effect of the nondimensional depletion length $\Lambda_d$ on the nondimensional prefactor $\eta_f$. 
Experimental $\Lambda_d$ (\ref{eq:Lambda_d}) follows from table \ref{tab:surfactant_values} and the experimental $\eta_f$ is as in Fig.~\ref{fig:exp_analysis}(b). The error bars are from the uncertainty of the physical values $k_a k_d^{-1}$ ($\pm 30\%$), $\Delta \Sigma_0$ ($\pm 1$ mN/m), $r_d$ ($\pm 2$ \textmu m), and $h_i$ ($\pm 5$ \textmu m).  
The theoretical  prediction (\ref{eq:front_prefac_prediction}) for $\Delta \Sigma_0 = 5$ mN/m is shown as a solid curve. The shaded region is the range of theoretical predictions accounting for the range $\Delta \Sigma_0 = 2.5-10$ mN/m. The theory uses the experimental values $(r_d, h_i, \Sigma, \mu, \rho) = (22~\text{\textmu m},\ 25~\text{\textmu m},\ 72~\text{mN/m},\ 0.9~\text{mPas},\ 997~\text{kg m}^{-3})$. The dotted line shows the theoretical asymptote (\ref{eq:front_asymptote}) for $\Lambda_d \ll 1$. There are no fitting parameters.
}
\end{figure}

Comparisons between the theoretical prediction (\ref{eq:front_prefac_prediction}) and the experimental data are shown in Fig.~\ref{fig:theory_comp}, where the front prefactor is plotted against $\Lambda_d$. There are no fitting parameters; all are from experimental measurements.  The experimental data are given by symbols and the theoretical prediction is given by a solid curve.
The dotted curve shows the asymptote of the theory (\ref{eq:front_asymptote}) for $\Lambda_d \ll 1$. There is excellent agreement, and the $\Lambda_d^{-1/4}$ scaling for the front growth is visible both experimentally and theoretically.

In conclusion, we have shown experimentally and theoretically that surfactant solubility enhances surfactant-induced film thinning due to the replenishment of surface surfactants from the bulk. The dynamics are dictated by a single nondimensional parameter $\Lambda_d$, which enhances the effective surface tension deficit (\ref{eq:eff_surf_tension}) with a $\Lambda_d^{-1/4}$ front-propagation scaling for $\Lambda_d \ll 1$.  Theoretically, such scaling implies that the minimum thickness is proportional to $\Lambda_d^{1/2}$ (see end matter), which explains the thin region seen for S8S in Fig.~\ref{fig:setup}(b). 
Our results suggest the front propagation velocity differs by roughly $4$ between S8S and S14S, implying a factor $4^2 = 16$ difference in the minimum thickness thinning rate. 
The order-of-magnitude difference in the thinning rate will significantly impact whether surfactant-induced thinning ultimately leads to rupture, 
such as by van der Waals forces \cite{Vaynblat02,Wee22,Wee24}, background flows \cite{Burton07,Fontelos18}, or entrained bubbles \cite{Lhuissier13,Lohse20,Dixit26}.
This is visible without a camera: spraying the Savart sheet with S$n$S, the sheet shows visibly greater disruption for small $n$, reflecting the likelihood of hole formation before the deposition pattern reaches the sheet edge. 

The essence of the enhancement mechanism is as follows. 
The region behind the surfactant front is, as a whole, the thinnest region of the sheet and  thins over time ($\textit{O}(t^{-1})$, see end matter), pushing (due to increased surface-area-to-volume ratio) surfactants onto the surface, and making finitely soluble surfactants effectively insoluble at late times.
The important parameter is the depletion length $k_a k_d^{-1}$ relative to the film thickness, i.e., $\Lambda_d$ (\ref{eq:Lambda_d}). 
By contrast, the surfactant-containing region does not thin over time as a whole for air-liquid-solid films or a deep liquid subphase. For the former, the film thickness approaches a steady profile in self-similar spatial coordinates, i.e., remaining $\textit{O}(t^0)$ \cite{Jensen92,Jensen93}, since the no-slip condition at the solid surface resists motion more than in air-liquid-air films. 

Future work could investigate the influence of the enhancement via solubility on the stability of liquid sheets in more complex systems, e.g., predicting surface bubble lifetimes. It will also be interesting to consider multiple-component surfactant-laden solutions. 
Technologically, observing the surfactant front growth gives information about the solution composition, with the advantage that only picolitres of the solution (the deposited drop) are required. Finally, the problem of soluble surfactant deposition allows for more analysis both numerically and theoretically, which will be published elsewhere \cite{Eshima26_long_paper}.

\section*{Acknowledgments}
\begin{acknowledgments}
Experimental data taken for this study (7 surfactants at three concentrations deposited 30 times leading to 630 instances) is available at \cite{Eshima26_letter_PDC}.

We thank Stefano Brizzolara, Sacha Escudier, Jonghyun Hwang and Sam Koblensky for technical help with the experimental setup and Baptiste Néel for discussion on the setup. 
We thank the anonymous reviewers whose suggestions helped to improve the manuscript.
This research was conducted under the NSF Grant No. CBET-2246791 to H.A.S.; the Princeton SEAS Innovation Fund and NSF Grant No. 2242512 to L.D.; and the Eli and Britt Harari Fellowship in the Department of Mechanical and
Aerospace Engineering, Princeton University to J.E.

\end{acknowledgments} 
\section*{End Matter}
Here, we expand on the details of the experimental modelling assumptions and show the derivation of (\ref{eq:r_f_eff}) from the similarity solution of insoluble surfactant deposition \cite{Eshima25_PRL, Eshima25_JFM}. 

\textit{Experimental modelling assumption details--- } First, we explain why stretching due to the background flow of the Savart sheet is assumed weak relative to the flow induced by the Marangoni stress.
The background flow is directed outward from the jet and is nearly uniform \cite{Marmottant00} with velocity $v_j \approx 4$ m s$^{-1}$, so a patch of radius $r_f$ located at distance $r_s$ from the jet center is stretched azimuthally (relative to the jet center) at velocity $\approx v_j r_f/r_s$. Integrating over time with $r_f$ proportional to $t^{1/2}$ gives deformation $\approx (2/3)v_j r_ft/r_s \lesssim 0.2 r_f$ for $t<5$ ms (taking $r_s \approx 6.5$ cm); the stretching is thus weak and the patterns seen in Fig.~\ref{fig:setup}(b) are approximately axisymmetric. Furthermore, we measure $r_f$ along the ray from the jet center to minimize background azimuthal stretching effects on our measurements. Specifically, since the focal plane of the camera lies along the ray from the jet center (see Fig.~\ref{fig:setup}(a)), $r_f$ is taken as half the horizontal chord intersecting the ellipse in the frame (see Fig.~\ref{fig:setup}(b)).

Second, we explain why the drop coalescence dynamics can be neglected.
The coalescence is estimated to occur on the inertiocapillary timescale $\sqrt{\rho r_d^3/\Sigma} \approx 0.01$ ms \cite{Kavehpour15} (where $\Sigma$ is the characteristic surface tension), consistent with experiments where coalescence is seen to complete within $\approx 0.1$ ms, far shorter than the timescale of the surfactant front growth (in our experiments, observable up to $\approx 5$ ms). In other words, $r_d$ in the experiment is small enough that the details of the drops do not need to be considered.

\textit{Derivation of the similarity solution---} For insoluble surfactant deposition with initial surfactant concentration $\Gamma_i$ (top-bottom symmetric) onto a flat sheet of thickness $h_i$ otherwise at rest, letting $N_{\Gamma}=2\pi \int_0^{\infty}\Gamma_i r dr$ and $\Gamma_m = \max \Gamma_i$, the characteristic radial lengthscale $\mathcal{L}$ and timescale $\mathcal{T}$ are given by 
\begin{subeqnarray}
\mathcal{L}&:=&\left(\frac{N_{\Gamma}}{\pi \Gamma_m}\right)^{1/2},\slabel{eq:L_theory_defn}\\
    \mathcal{T} &:=&\left(\frac{\rho h_i \mathcal{L}^2}{\Delta \Sigma}\right)^{1/2},
\end{subeqnarray}
where the characteristic surface tension deficit $\Delta \Sigma := -\Gamma_m \left.\frac{d\sigma}{d \Gamma}\right|_{\Gamma = 0}$. Let $\mu$ and $\Sigma$ be the dynamic viscosity and surface tension of pure DI water respectively. There are two non-dimensional numbers
\begin{subeqnarray}
    \mathcal{M} &:=& \frac{\Delta \Sigma \mathcal{L}^2}{h_i^2\Sigma} = \frac{N_\Gamma\left(-\left.\frac{d\sigma}{d\Gamma}\right|_{\Gamma = 0}\right)}{\pi h_i^2\Sigma},\\ \textit{Re} &:=& \left(\frac{\rho \Delta \Sigma \mathcal{L}^2}{h_i\mu^2}\right)^{1/2}=\left(\frac{\rho N_\Gamma\left(-\left.\frac{d\sigma}{d\Gamma}\right|_{\Gamma = 0}\right)}{h_i\mu^2 \pi}\right)^{1/2},\label{eq:M_Re_insoluble}
\end{subeqnarray}
where the Marangoni number $\mathcal{M}$ is the ratio of Marangoni to capillary stress and the Reynolds number $\textit{Re}$ is the ratio of inertia to viscous extensional stress. 

Then, assuming a thin-film limit $h_i/\mathcal{L}\ll 1$,
the following nondimensional equations are found for radial velocity $\tilde{u}(\tilde{r},\tilde{t})$, thickness $\tilde{h}(\tilde{r},\tilde{t})$, and surface surfactant concentration $\tilde{\Gamma}(\tilde{r},\tilde{t})$, with nondimensional variables defined by $\tilde{r} = r/\mathcal{L}, \tilde{t} = t/\mathcal{T}, \tilde{u} = u \mathcal{T} /\mathcal{L}, \tilde{h} = h/ h_i, \tilde{\Gamma}=\Gamma/\Gamma_m$:
\begin{subeqnarray}
    \frac{\partial \tilde{u}}{\partial \tilde{t}}+ \tilde{u}\frac{\partial \tilde{u}}{\partial \tilde{r}} &=& -\frac{2}{\tilde{h}}\frac{\partial \tilde{\Gamma}}{\partial \tilde{r}} + \frac{1}{2 \mathcal{M}}\frac{\partial }{\partial \tilde{r}}\left(\frac{1}{\tilde{r}}\frac{\partial}{\partial \tilde{r}}\left(\tilde{r}\frac{\partial \tilde{h}}{\partial \tilde{r}}\right)\right)\nonumber \\
    &+&
   \frac{4}{\textit{Re}} \frac{1}{\tilde{h}}\left(\frac{\partial}{\partial \tilde{r}}\left(\frac{\tilde{h}}{\tilde{r}}\frac{\partial }{\partial \tilde{r}}(\tilde{r} \tilde{u})\right) - \frac{1}{2} \frac{\tilde{u}}{\tilde{r}}\frac{\partial \tilde{h}}{\partial \tilde{r}}\right),\slabel{eq:tf_mom}\\
    \frac{\partial \tilde{h}}{\partial \tilde{t}}&=&-\frac{1}{\tilde{r}}\frac{\partial }{\partial \tilde{r}}\left(\tilde{r} \tilde{u} \tilde{h}\right),\slabel{eq:tf_mass}\\
    \frac{\partial \tilde{\Gamma}}{\partial \tilde{t}}&=&-\frac{1}{\tilde{r}}\frac{\partial }{\partial \tilde{r}}\left(\tilde{r} \tilde{u} \tilde{\Gamma}\right).\label{eq:tf_flow_total}
\end{subeqnarray}
When $\mathcal{M}^{-1}=\textit{Re}^{-1}=0$, system (\ref{eq:tf_flow_total}) is mathematically the 2D axisymmetric compressible gas equations upon replacing $-2\tilde{\Gamma}$ with pressure, and $\tilde{h}$ with the density (and pressure/density$^{\gamma}=$ constant for $\gamma=1$) \cite{Chomaz01, Kim17}. 

The late-time $\tilde{t} = t/\mathcal{T} \gg 1$ similarity solutions are found by matching the three regions as shown in Fig.~\ref{fig:setup}(c).

Below, we briefly summarise scaling arguments in \cite{Eshima25_PRL,Eshima25_JFM}. Subscripts denote the region under discussion. Let $\Delta \tilde{r}$ be the characteristic width of a given region. Since the film is undisturbed at far field, $\tilde{h}_{\text{III}} \sim 1$. Since $\tilde{u}_{\text{III}}\sim \Delta \tilde{r}_{\text{III}} \tilde{t}^{-1}$, balancing inertia $\tilde{u}_{\text{III}}\tilde{t}^{-1}$ with capillary $\tilde{h}_{\text{III}}(\Delta \tilde{r}_{\text{III}})^{-3}$ and/or viscous extensional stress $\tilde{u}_{\text{III}}(\Delta \tilde{r}_{\text{III}})^{-2}$ gives $(\tilde{u}_{\text{III}},\tilde{h}_{\text{III}},\Delta \tilde{r}_{\text{III}}) \sim (\tilde{t}^{-1/2},\tilde{t}^0,\tilde{t}^{1/2})$. Upon integrating across region II where there is a balance between Marangoni and capillary stress, which can be thought of as a jump region, $\tilde{\Gamma}_{\text{I}}\sim \tilde{h}_{\text{III}}^2 (\Delta \tilde{r}_{\text{III}})^{-2} \sim \tilde{t}^{-1}$. Then by global conservation of surfactants, $\Delta \tilde{r}_{\text{I}}\sim \tilde{t}^{1/2}$. Hence $\tilde{u}_{\text{I}} \sim \tilde{t}^{-1/2}$ and using that $\tilde{\Gamma}_{\text{I}}/\tilde{h}_{\text{I}}$ is conserved in Lagrangian coordinates \cite{Chomaz01}, it follows that $\tilde{h}_{\text{I}} \sim \tilde{t}^{-1}$. Thus, $(\tilde{u}_{\text{I}},\tilde{h}_{\text{I}},\tilde{\Gamma}_{\text{I}},\Delta \tilde{r}_{\text{I}})\sim (\tilde{t}^{-\frac{1}{2}},\tilde{t}^{-1},\tilde{t}^{-1},\tilde{t}^{\frac{1}{2}})$.

In particular, the surfactant front and minimum thickness at late times $\tilde{t} = t/\mathcal{T} \gg 1$ are given respectively by \cite{Eshima25_PRL, Eshima25_JFM}
\begin{equation}
    \tilde{r}_f = \eta_f^{\text{insol}} \tilde{t}^{1/2},~\tilde{h}_{\text{min}}=(\eta_f^{\text{insol}})^{-2}\tilde{t}^{-1}, \label{eq:insoluble_simil_soln}
\end{equation}
where $\tilde{r}_f = r_f/\mathcal{L}$, $\tilde{h}_{\text{min}}=h_{\text{min}}/h_i$, and $\eta_f^{\text{insol}} = \eta_f^{\text{insol}}(\mathcal{M},\textit{Re})$ is a prefactor, set by $\mathcal{M}$ and $\textit{Re}$ that can be found from a shooting method for the similarity solution ODEs \cite{Eshima25_PRL, Eshima25_JFM}.  

In order to compare against experiments tracking the surfactant front, we wish to convert (\ref{eq:insoluble_simil_soln}) back to dimensional variables. Such a step involves the factor
\begin{equation}
    \mathcal{L} \mathcal{T}^{-1/2}=\left(\frac{\Delta \Sigma \mathcal{L}^2}{\rho h_i}\right)^{1/4} = \left(\frac{N_\Gamma\left(-\left.\frac{d\sigma}{d\Gamma}\right|_{\Gamma = 0}\right)}{\pi\rho h_i}\right)^{1/4}.\label{eq:scale}
\end{equation}

As discussed in the main text the total amount of surfactants deposited in the experiment is $(4/3)\pi r_d^3 c_0$. Then, the top-bottom symmetric insoluble surfactant deposition with the same amount of surfactants satisfies
\begin{equation}
    2N_\Gamma = \frac{4}{3}\pi r_d^{3}c_0\label{eq:transform_mass}.
\end{equation}
Substituting (\ref{eq:transform_mass}, \ref{eq:static_surface_tension}) into (\ref{eq:M_Re_insoluble}, \ref{eq:insoluble_simil_soln}, \ref{eq:scale}) gives the expression
\begin{equation}
    r_f =\eta_f^{\text{insol}} (2\Lambda_d)^{-1/4}\left(\frac{4 r_d^3\Delta \Sigma_0}{3\rho h_i^2}\right)^{1/4}t^{1/2},\label{eq:r_f_prediction}
\end{equation}
as desired, where $\eta_f^{\text{insol}} = \eta_f^{\text{insol}}(\mathcal{M},\textit{Re})$ with $\mathcal{M}$ and $\textit{Re}$ given by 
\begin{subeqnarray}
    \mathcal{M} &=& (2\Lambda_d)^{-1}\frac{4r_d^3\Delta \Sigma_0}{3h_i^3\Sigma}\\
    \textit{Re}&=&(2\Lambda_d)^{-1/2}\left(\frac{4\rho \Delta \Sigma_0 r_d^3}{3h_i^2\mu^2}\right)^{\frac{1}{2}}.
\label{eq:M_Re_fin_soluble}
\end{subeqnarray}
It is also observed that a factor $\Lambda_d^{1/2}$ for $h_{\text{min}}$ arises since $h_{\text{min}}=(\eta_f^{\text{insol}})^{-2}(2\Lambda_d)^{1/2}[\rho h_i^4\mathcal{L}^4/([4/3] r_d^3 \Delta \Sigma_0)]^{1/2}t^{-1}$. Note that the predictions (\ref{eq:r_f_prediction},\ref{eq:M_Re_fin_soluble}) vanish in the insoluble limit $\Lambda_d \rightarrow \infty$ since the total amount of surfactants deposited $(4/3) \pi r_d^3 c_0$ vanishes as $c_0\rightarrow 0$ (for fixed $\Delta \Sigma_0$, larger $k_a k_d^{-1}$ leads to smaller $c_0$, see (\ref{eq:static_surface_tension})). 

A technical remark is that a particular definition of lengthscale $\mathcal{L} = [(4/3)r_d^3 h_i^{-1}]^{1/2}$ was chosen to nondimensionalise the experimental data and to define $\eta_f$ (\ref{eq:front_prefac}, \ref{eq:explicit_L}), which assumes that $\Gamma_m = h_i c_0/2$ (see the definition (\ref{eq:L_theory_defn})). However, the theoretical prediction (\ref{eq:r_f_prediction}, \ref{eq:M_Re_fin_soluble}) is independent of $\Gamma_m$ and therefore does not require assumptions about the exact value of $\Gamma_m$. 

\bibliography{mybib}

\end{document}